\documentclass[aps,prd,amsmath,amsfonts,a4paper,11pt,reprint,preprintnumbers,twocolumn,nofootinbib]{revtex4-1}
\usepackage{hyperref}
\usepackage{natbib}
\usepackage{amsmath}
\usepackage{amsfonts}
\usepackage{amssymb}
\usepackage{color}
\usepackage{bm}

\usepackage{graphicx}
\usepackage{subfigure}

\renewcommand{\d}{\mathrm{d}}

\newcommand{\vect}[1]{\bm{\mathrm{{#1}}}}
\newcommand{\e}[1]{\mathrm{e}^{{#1}}}

\renewcommand{\leq}{\leqslant}

\DeclareMathOperator{\Or}{O}

\newcommand{\ksmooth}{k_s}
\newcommand{\fNL}{f_{\mathrm{NL}}}
\newcommand{\etal}{et al.}
\newcommand{\Mp}{M_{\mathrm{P}}}

\newcommand{\para}[1]{\par\vspace{2mm}\noindent\textbf{{#1}.}}

\begin{document}

%======================================%
%<<<<<<<<<<<< TITLE PAGE >>>>>>>>>>>>>>%
%======================================%

\title{Transport equations for the inflationary spectral index}
\author{Mafalda Dias}
\affiliation{Astronomy Centre, University of Sussex, Brighton BN1 9QH,
  United Kingdom}
\author{David Seery}
\affiliation{Astronomy Centre, University of Sussex, Brighton BN1 9QH,
  United Kingdom}
\date{\today}

%======================================%
%<<<<<<<<<<<<< ABSTRACT >>>>>>>>>>>>>>>%
%======================================%

\begin{abstract}

We present a simple and efficient
method to compute the
superhorizon evolution of the spectral index
in multi-field inflationary models, using transport equation techniques.
We illustrate the evolution
of $n_s$ with time for various interesting potentials.

\end{abstract}

\maketitle

%======================================%
%<<<<<<<<<<<<<< ARTICLE >>>>>>>>>>>>>>>%
%======================================%

\section{Introduction}

The imminent arrival of precise microwave background data
from \emph{Planck} makes this a promising time to put
constraints on cosmological models, and in particular
the inflationary paradigm.
To do so, we need accurate calculations of the observables
produced by an inflationary phase,
of which we expect
the spectral index of the density perturbation, $n_s$,
to be among the most precisely measured.
\emph{Planck} may
determine $n_s$ with an error of perhaps a few times
$10^{-3}$~\cite{Colombo:2008ta}.

Inflationary models motivated by high-energy physics
often invoke many light scalar fields.
The resulting multi-field dynamics can
cause the curvature perturbation synthesized at horizon crossing
to evolve, as power is transferred from decaying isocurvature
modes. During this time the spectral index will typically vary.
Accurately tracking its trajectory
is an important challenge,
and
various techniques to carry out the computation have been proposed.

Statistical properties of the inflationary density perturbations
can be calculated using the separate universe
approximation.
A simple way to implement this approximation
is to Taylor expand in powers of the initial
conditions~\cite{Starobinsky:1986fxa, *Lyth:2005fi}.
When applied to the
curvature perturbation
this is
sometimes called the ``$\delta N$ formalism''. Long ago,
Sasaki \& Stewart used this method to provide a very general formula
for the spectral index~\cite{Sasaki:1995aw},
valid for an arbitrary choice of field space.
This formula has proved entirely satisfactory for analytic calculations.
But our ambition to constrain increasingly complex models
means that numerical work is often needed, particularly for
the sophisticated examples motivated by high-energy
theory~\cite{Frazer:2011tg, Agarwal:2011wm}.
For these cases, a \emph{direct} implementation of the $\delta N$
formula is less attractive, because it is not framed in terms
of ordinary differential equations but rather ``variational''
derivatives which
require numerical integration
of high accuracy. For a discussion, see Ref.~\cite{Mulryne:2010rp}.

An alternative approach
was developed
by Gordon~{\etal} and by
Nibbelink \& van Tent~\cite{Gordon:2000hv, *GrootNibbelink:2001qt}.
Their formalism is equivalent to the Taylor expansion
method, although more complicated because it requires decomposition
in terms of a Frenet basis for the inflationary trajectory in field
space.
Recently, Peterson \& Tegmark used this method to compute the spectral
index
\cite{Peterson:2010np, *Peterson:2010mv, *Peterson:2011yt}.
But
it would be desirable to have a simple system of ordinary differential
equations,
expressed directly in terms of the potential,
which do not invoke these complications.
In this short paper we provide such a system.

To achieve this,
we explore the evolution of
correlation functions in a
multi-field framework using transport techniques
\cite{Mulryne:2009kh, Mulryne:2010rp}.
We obtain a transport equation which propagates
the spectral index from its value at horizon crossing
up to the end of inflation.
The result is technically equivalent to those of
Stewart \& Sasaki and Peterson \& Tegmark, but
we believe it has advantages of simplicity and numerical implementation.
In addition, it is easy to study the entire history of the spectral
index during inflation, which can be correlated with other
observables such as the amplitude of the local-mode bispectrum.
We give an example in \S\ref{sec:examples}.

\section{Superhorizon evolution with transport equations}
\label{sec:transport}

The ``transport'' approach
is another implementation of the separate universe
approximation
\cite{Mulryne:2009kh, Mulryne:2010rp, Starobinsky:1986fxa, *Lyth:2005fi}.
We restrict to canonical kinetic terms but allow an arbitrary number of
fields, and set $c = \hbar = 1$.
The reduced Planck mass is $\Mp^2 = (8 \pi G)^{-1}$.
The species of light scalar fields are indexed with Greek labels,
and we make use of the slow-roll approximation throughout.

In the standard picture, the fluctuations associated with
increasing wavenumbers
pass sequentially
outside the horizon.
After a few e-folds, the decaying mode of each fluctuation has died away
and it is treated as a classical
object~\cite{Lyth:1984gv, *Polarski:1995jg, *Lyth:2006qz}
(although a detailed understanding of the quantum-to-classical transition
is still lacking).
To determine the
subsequent evolution we invoke the separate universe approximation.
Consider any large spatial region, smoothed on a scale
larger than the horizon.
The separate universe approximation
asserts that, at any position, the coarse-grained fields evolve
as they would in an unperturbed universe, up to gradient-suppressed
corrections.

The observable predictions of any inflationary model are not the fields
themselves, but their correlation functions.
In Refs.~\cite{Mulryne:2009kh, Mulryne:2010rp} these were studied
using the distribution of coarse-grained fields
obtained after smoothing on the scale $\ksmooth$.
To leading order, inflation predicts that this probability
distribution is Gaussian
and can therefore be characterized by
the expectation value $\langle \phi_\alpha \rangle \equiv
\Phi_\alpha$
and variance $\langle \delta \phi_\alpha \delta \phi_\beta \rangle \equiv
\Sigma_{\alpha\beta}$
of the coarse-grained fields,
where $\delta \phi_\alpha \equiv \phi_\alpha - \Phi_\alpha$.
These one- and two-point functions evolve according to%
\footnote{These equations were given for an inflationary model
with an arbitrary number of fields
in Refs.~\cite{Mulryne:2009kh, Mulryne:2010rp}.
In the single-field case they appear in Ref.~\cite{citeulike:1400625}.

After extinction of the decaying mode
we are neglecting any further quantum effects.
In principle, the probability distribution of the coarse-grained fields
should be inherited from the generating function of correlation functions,
ie., the effective action.
In a closed-time-path formulation there is no effective action in
the usual sense, because its evolution
(and that of the correlation functions) becomes stochastic.
See, for example, Ref.~\cite{Calzetta:1996sy}.
Over sufficiently long times this stochasticity
cannot be neglected, and Eqs.~\eqref{eq:centroid}--\eqref{eq:variance}
should be supplemented by Langevin terms, as originally
proposed by Starobinsky
\cite{Starobinsky:1986fx, *Starobinsky:1994bd,
*Salopek:1990re, *Salopek:1990jq,
*Seery:2009hs}.
In the full quantum theory, one would expect the evolution of
correlation functions to be determined by a Schwinger--Dyson
hierarchy. The transition to a Boltzmann--Langevin hierarchy
was studied
by Calzetta \& Hu~\cite{Calzetta:1999xh}.}
\begin{subequations}
\begin{align}
	\label{eq:centroid}
	\Phi'_{\alpha} & =
		\phi'_{\alpha}
		+ \frac{1}{2}u_{\alpha\lambda\mu}\Sigma_{\lambda\mu} + \cdots \\
	\label{eq:variance}
	\Sigma'_{\alpha\beta} & =
		u_{\alpha \lambda}\Sigma_{\lambda \beta}
		+ u_{\beta \lambda}\Sigma_{\lambda \alpha} + \cdots ,
\end{align}
\end{subequations}
where we have uniformly omitted correlation functions of third-order
or higher.
The $u$-tensors are related to the field velocity
$\phi'_\alpha$ and can be obtained using perturbation theory;
for details, see Refs.~\cite{Mulryne:2009kh, Mulryne:2010rp}.
In general, each $n$-point function will be sourced by all other
correlation functions, and for computational purposes the hierarchy must be
truncated.
In~\eqref{eq:centroid}, the truncation
of three-point functions and above leaves a ``loop correction''
to the one-point function, which is negligible in practical examples
\cite{Lyth:2006qz, Boubekeur:2005fj, *Lyth:2007jh, *Seery:2007wf,
*Bartolo:2007ti, *Seery:2010kh}.
Eqs~\eqref{eq:centroid}--\eqref{eq:variance}
should be supplemented by boundary conditions obtained from matching
to the known $n$-point functions computed
perturbatively at horizon-crossing.
We will encounter an explicit example
in \S\ref{sec:transport-ns} below.

For comparison with experiment
we require the correlation functions of the curvature
perturbation, which seeds the observable density fluctuation
\cite{Bardeen:1980kt, *Bardeen:1983qw, *Wands:2000dp, *Maldacena:2002vr}.
Eqs.~\eqref{eq:centroid}--\eqref{eq:variance} apply for any choice of
correlation functions, but they are simplest to solve when
our choice makes the system autonomous.
% ie., the u_i do not depend on the time variable
A suitable choice is the
$n$-point functions of the canonically normalized scalar fields
whose potential energy supports the inflationary era.
The correlation functions of the curvature perturbation, $\zeta$,
can be obtained from these by a gauge transformation.
At fixed cosmic time $t = t_c$, this transformation can be written
\begin{equation}
	\label{eq:zeta}
	\zeta =
		N_{,\alpha}\delta\phi_{\alpha}
		+ \frac{1}{2}N_{,\alpha\beta}
			(\delta\phi_{\alpha}\delta\phi_{\beta}
				- \langle\delta\phi_{\alpha}\delta\phi_{\beta}
			\rangle)
		+ \cdots ,
\end{equation}
where $N = \int H \, \d t$ measures the number of e-foldings of inflation
and
the omitted terms are higher-order in powers of the small
fluctuations $\delta \phi_\alpha$.
All quantities in~\eqref{eq:zeta}
depend only on data local to the hypersurface
$t = t_c$.
Unless loop corrections are being retained,
it is only necessary to keep contributions to the 2-point function from
$N_{,\alpha}$,
dropping $N_{,\alpha\beta}$ and any higher Taylor
coefficients \cite{Lyth:2007jh, *Seery:2007wf}.
Also, $N_{,\alpha}$
can be written in terms of %derivatives of
the inflationary
potential, $V(\phi)$, using
\begin{equation}
	N_{,\alpha} =\sum_{\beta}
		\frac{V}{V,_{\alpha}}
		\frac{V,_{\alpha}^{2}}{V,_{\beta}^2}
	\label{eq:gauge-ni}
\end{equation}
Eq.~\eqref{eq:gauge-ni} applies for an arbitrary $V$
\cite{Mulryne:2009kh, Mulryne:2010rp, Malik:2008im}
but invokes the slow-roll approximation.
(The summation convention is suspended for~\eqref{eq:gauge-ni},
but used elsewhere in this paper.)
The two-point function of $\zeta$ can be calculated
using \cite{Starobinsky:1986fxa, *Lyth:2005fi, Sasaki:1995aw}
\begin{equation}
\langle \zeta \zeta \rangle
= N_{,\alpha} N_{,\beta} \Sigma_{\alpha\beta}.
\end{equation}

Eqs.~\eqref{eq:centroid}--\eqref{eq:variance}
and~\eqref{eq:zeta}--\eqref{eq:gauge-ni}
give a practical algorithm with which to study evolution of the power
spectrum on the smoothing scale $\ksmooth$.
By carrying out the computation for a few closely spaced $\ksmooth$
one could numerically study the $\ksmooth$ dependence of the result.
But this approach would be undesirable,
requiring~\eqref{eq:centroid}--\eqref{eq:variance} to be integrated
several times for each model of interest.
For example,
if the resulting inflationary observables are to be used
in drawing a sample from the space of models
(perhaps as input to a code such as CosmoMC or MultiNest,
which may require
exponentially many samples),
then the resulting time penalty can become appreciable.
%In some cases, constructing a satisfactory numerical approximation
%to the $\ksmooth$-derivative may also require care.

\section{Transport of the spectral index}
\label{sec:transport-ns}

To do better, we return to Eq.~\eqref{eq:variance} and use it to obtain
a transport equation for the scale dependence directly.
We write
\begin{subequations}
\begin{align}
	\langle\zeta(\vect{k}_1)\zeta(\vect{k}_2)\rangle
		& = (2 \pi)^{3} \delta^{3}(\vect{k}_{1} + \vect{k}_{2})
		\frac{P(k_{1})}{2 k_1^3}
	\\
	\langle
		\delta \phi_\alpha (\vect{k}_1) \delta \phi_\beta (\vect{k}_2)
	\rangle
		& = (2 \pi)^3 \delta^{3}(\vect{k}_1 + \vect{k}_2)
		\frac{\Sigma_{\alpha \beta}(k_1)}{2 k_1^3} .
	\label{eq:field-twopf}
\end{align}
\end{subequations}
Up to a normalization,
the function $P$ is the power spectrum of the curvature perturbation,
and can be computed using $P = N_{,\alpha} N_{,\beta}
\Sigma_{\alpha\beta}$.
The $k$-label
argument of $P$ and $\Sigma_{\alpha\beta}$
should be associated with the smoothing scale $\ksmooth$
discussed in \S\ref{sec:transport}.
To study the variation of $P$
with $\ksmooth$ at equal times, it is conventional to use the
parametrization
\begin{equation}
	\label{eq:taylor}
	P(\ksmooth) \simeq
		P_*(\ksmooth)
		\left( 1 + (n_{s}-1)_* \ln{\frac{\ksmooth}{k_{*}}} + \cdots \right)
\end{equation}
where the spectral index $n_s$ satisfies
\begin{equation}
	(n_{s}-1)_* \equiv
		\left.
			\frac{\d\ln{P(\ksmooth)}}{\d \ln{\ksmooth}}
		\right|_{\ksmooth = k_*} .
\end{equation}
The ``pivot'' scale $k_\ast$ is arbitrary and can be chosen for convenience.
It is analogous to the arbitrary renormalization scale in
scattering calculations.
Observable quantities would be independent of our choice
were all terms
in the expansion~\eqref{eq:taylor}
to be kept.
In practice, after truncating at finite order
it should be chosen approximately equal to the scale of interest,
$\ksmooth$,
to avoid an unwanted large logarithm.

The derivative with respect to $\ln{\ksmooth}$
is to be evaluated at equal times.
Directly differentiating the 2-point function of $\zeta$, we obtain
\begin{equation}
	n_{s}-1 =
		\frac{N_{,\alpha}N_{,\beta}}{P}
		\frac{\d \Sigma_{\alpha\beta}}{\d \ln{\ksmooth}}
		= \frac{N_{,\alpha}N_{,\beta} n_{\alpha\beta}}
			{N_{,\lambda} N_{,\mu} \Sigma_{\lambda\mu}} ,
	\label{eq:ns}
\end{equation}
where we have introduced the matrix $n_{\alpha\beta}
\equiv \d \Sigma_{\alpha\beta} / \d \ln k$,
which measures scale dependence of $\Sigma_{\alpha\beta}$.
Note that the gauge-transformation factors $N_{,\alpha}$
are $\ksmooth$-independent. They depend only on the
typical trajectory followed by the coarse-grained fields,
and the time of evaluation. The same is true for $u_{\alpha\beta}$,
meaning that it is simple to obtain an evolution equation for
$n_{\alpha\beta}$,
\begin{equation}
	\label{evolns}
	\frac{\d n_{\alpha\beta}}{\d N} =
		\frac{\d}{\d\ln k}
		\frac{\d \Sigma_{\alpha\beta}}{\d N} =
			u_{\alpha\lambda}n_{\lambda\beta}
			+ u_{\beta\lambda}n_{\lambda\alpha} .
\end{equation}
This is of precisely the same form as the general
transport equation~\eqref{eq:variance}.
Its solution will differ from $\Sigma_{\alpha\beta}$
through a different choice of boundary conditions.

Eq.~\eqref{evolns} is sufficient to determine
the evolution of $n_{s}$ from horizon crossing up to the end of inflation.
It is only necessary to specify the initial value of $n_{\alpha\beta}$.
To leading order in the slow-roll
approximation, the fluctuations of canonically normalized scalar fields
are uncoupled at horizon crossing,
making $\Sigma_{\alpha\beta}$ proportional to a Kronecker-$\delta$.
With the normalization of~\eqref{eq:field-twopf},
we have
$\Sigma_{\alpha\beta}|_\ast = H_\ast^2 \delta_{\alpha\beta}$.
Consider the pivot scale $k_\ast$ and a nearby shorter mode with wavenumber
$k = k_\ast( 1 + \delta \ln k )$
and $\delta \ln k > 0$.
When their decaying modes are lost, as described above, the
fluctuations associated with these wavenumbers settle down
to classical perturbations with amplitudes which differ
by \cite{Liddle:1992wi}
\begin{equation}
	\delta \Sigma_{\alpha\beta}|_{\ast}
		= - 2 \epsilon_*
			\Sigma_{\alpha\beta}|_\ast
			\; \delta \ln k .
\end{equation}
These two modes cross the horizon at slightly different times,
and therefore
when compared at the \emph{same} time
the longer mode $k_*$ experiences slightly more evolution.
It follows that 
there is an extra displacement $\delta \Sigma_{\alpha\beta}
\approx - (\d \Sigma_{\alpha\beta} / \d N)
\delta \ln k$, because $\delta \ln k$ measures the number of e-folds
which elapse between horizon exit of the two modes.
We conclude
\begin{equation}
	n_{\alpha\beta}|_\ast =
		- 2 ( \epsilon_{*} \delta_{\alpha\beta} +
			u_{\alpha\beta*} ) H^{2}_*.
	\label{eq:n-matrix-ic}
\end{equation}
Note that $u_{\alpha\beta}$ is symmetric when time is measured
in e-folds, $N$.

Eqs.~\eqref{eq:ns}--\eqref{eq:n-matrix-ic}
are all that is required to compute the evolution of the spectral index.
In the next section
we present some illustrative examples,
drawn for simplicity from the class of 2-field models.
However, we emphasize that this method is valid for any number of
scalar fields.

\begin{figure}
\centering
\includegraphics[width=6.2cm]{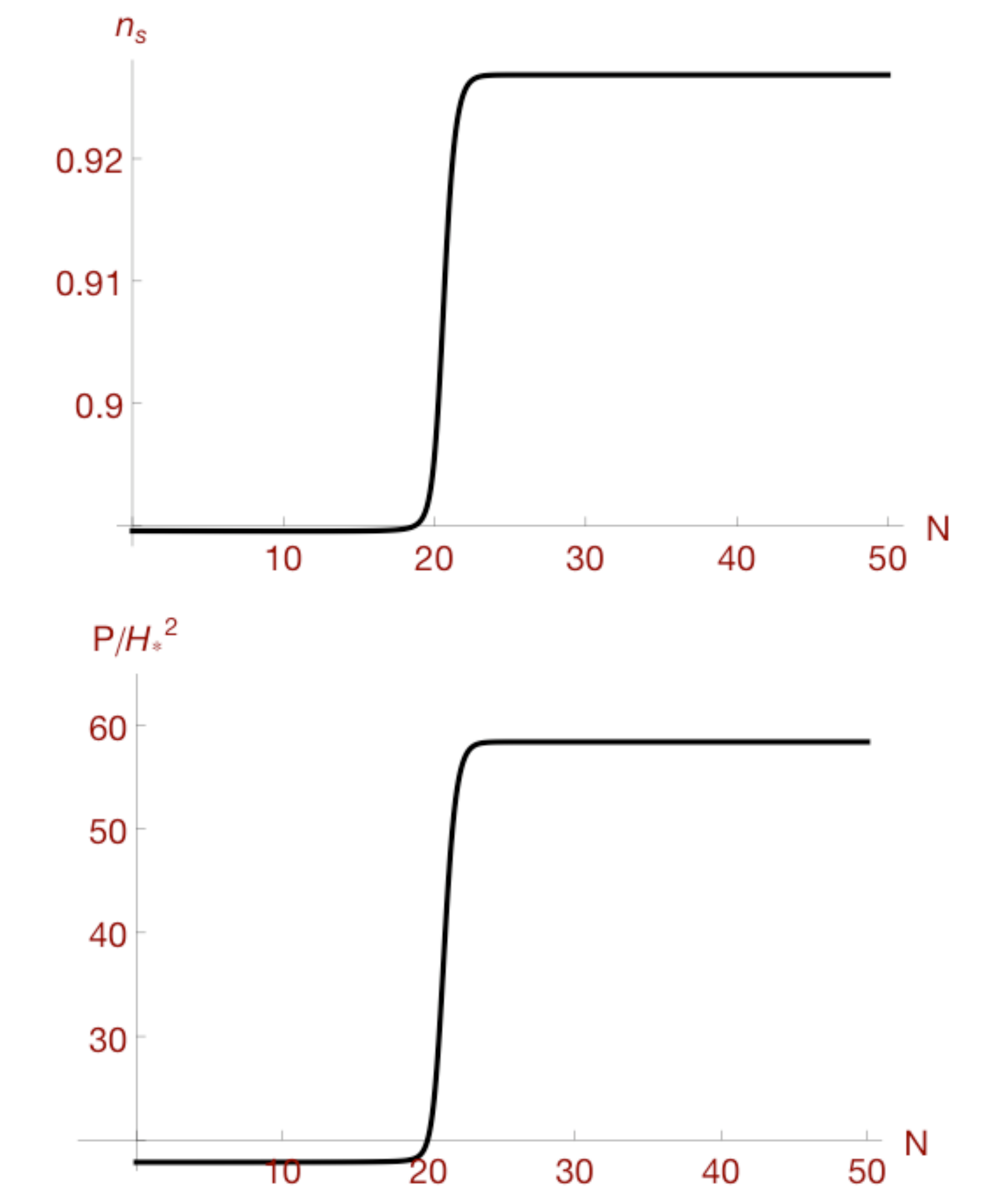}
\caption{Superhorizon evolution of the spectral index (top) and power spectrum of curvature perturbations (bottom) for the double quadratic potential.}
\label{doublequa}
\end{figure}

\begin{figure}
\centering
\includegraphics[width=6cm]{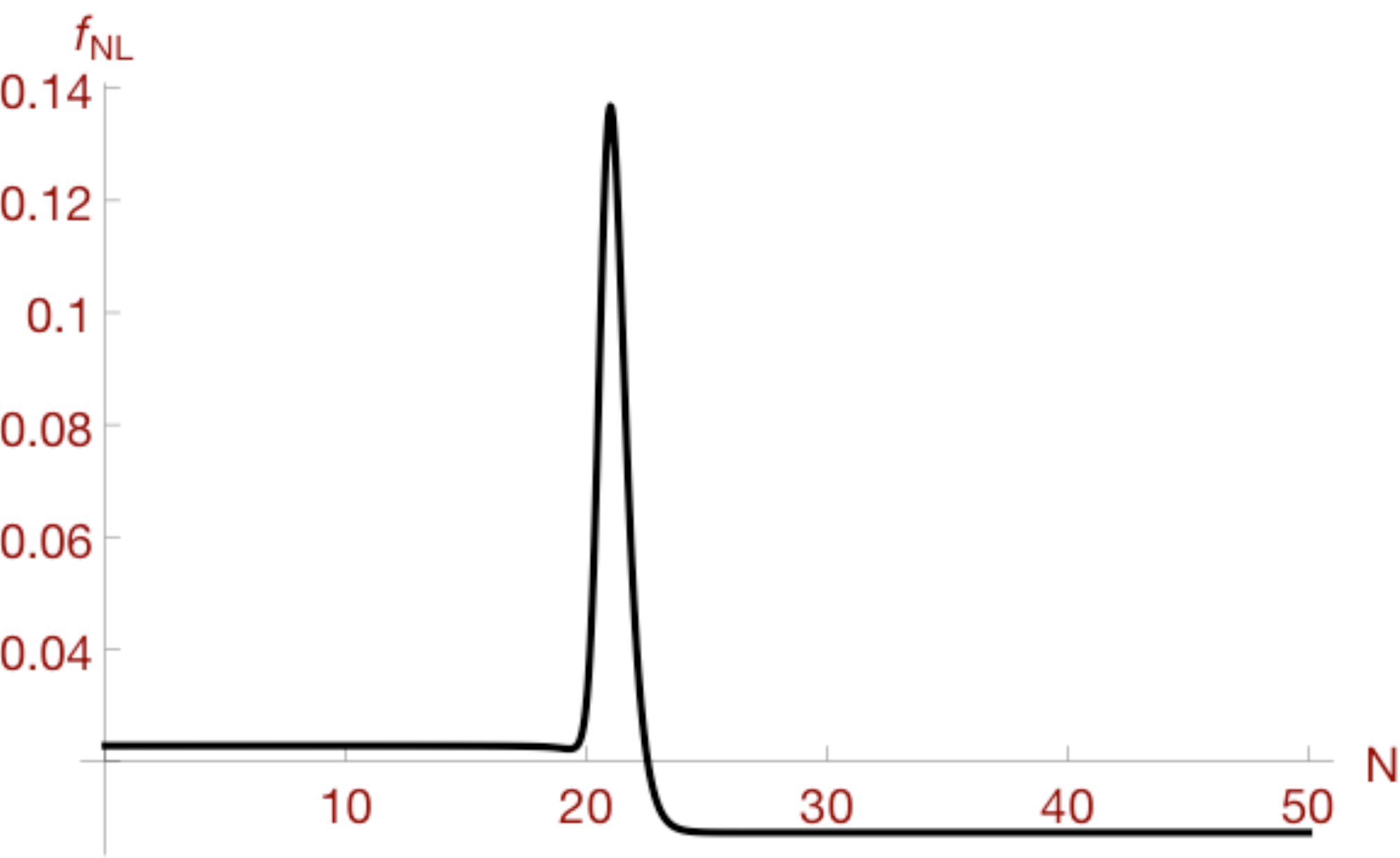}
\caption{Superhorizon evolution of the parameter $f_{NL}$ for the double quadratic potential.}
\label{fnl_dq}
\end{figure}

\section{Examples}
\label{sec:examples}
\para{Double quadratic inflation}
Our first example is the well-studied model of double quadratic
inflation~\cite{Gordon:2000hv, Langlois:1999dw,
*Rigopoulos:2005ae, *Rigopoulos:2005us, *Vernizzi:2006ve}
\begin{equation}
	V = \frac{1}{2} m_{\phi}^{2} \phi^2 + \frac{1}{2} m_{\chi}^{2} \chi^2
\end{equation}
where the mass ratio
$m_{\phi}/m_{\chi} = 9$ and $m_{\chi}= 10^{-5} \Mp$.
We begin the calculation
50 e-folds before the end of inflation,
with
$\phi=8.2 \Mp$ and $\chi=12.9 \Mp$.
In this model there is a single turn in field space.
The turn is
associated with a `spike' in the amplitude of the local-mode
bispectrum, typically measured by the parameter $\fNL$;
see Refs.~\cite{Vernizzi:2006ve, Mulryne:2009kh, Mulryne:2010rp}.
The turn is also associated with a step in the amplitude of the power
spectrum, $P$,
and a similar step in the spectral index.
In Figs.~\ref{doublequa} and~\ref{fnl_dq} these behaviours can be compared.
%The spike and the step clearly correlate with each other.
%Correlations between observables of this kind will be important
%when using \emph{Planck} data to discriminate between models,
%but a detailed study is beyond this scope of this paper.
We leave a careful analysis for future work.

\begin{figure}
\centering
\includegraphics[width=6.2cm]{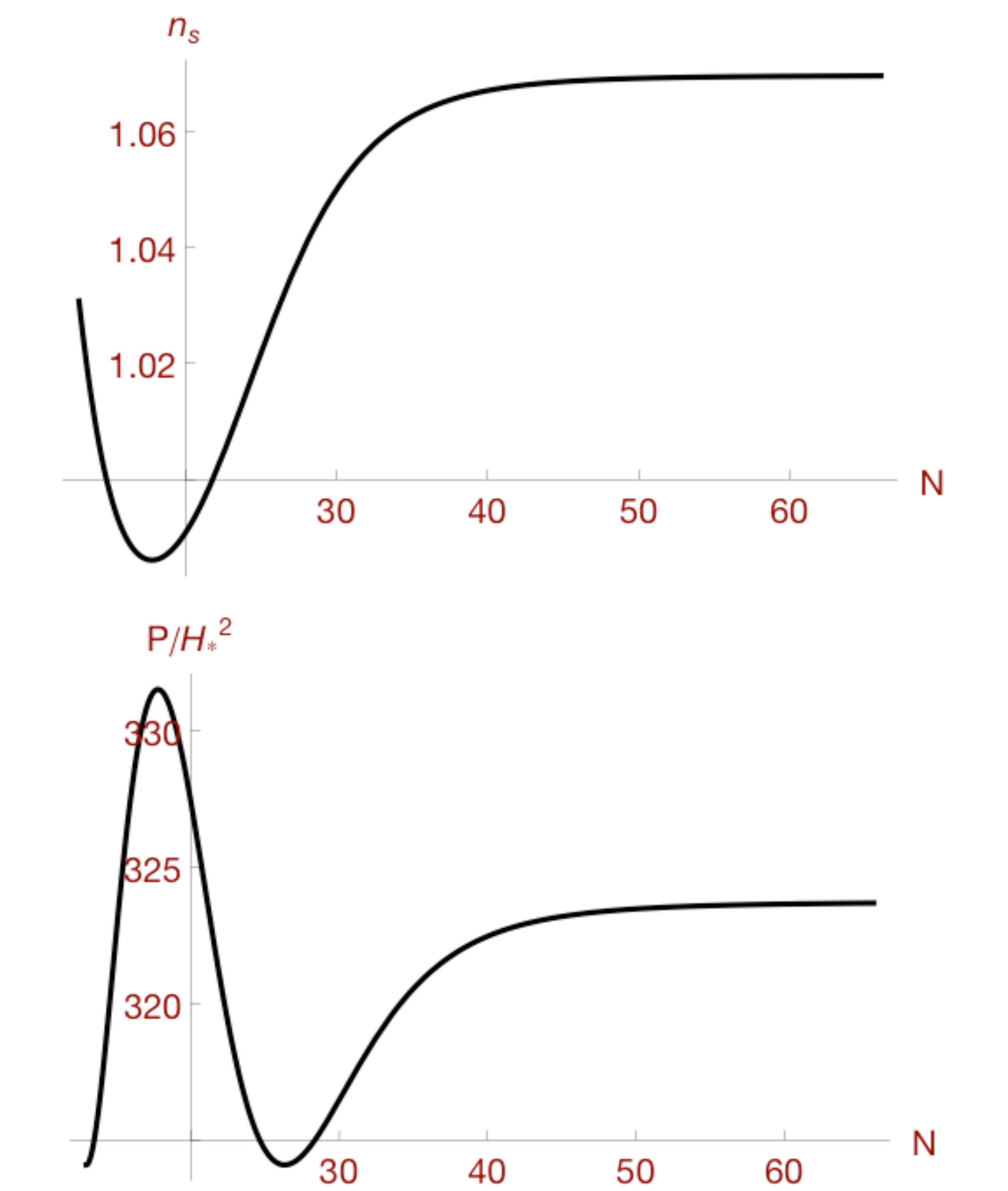}
\caption{Superhorizon evolution of the spectral index (top) and power spectrum of curvature perturbations (bottom) for a particular realization of a randomly generated potential.}
\label{nonmono}
\end{figure}

\para{Random potentials: nonmonotonic features}
In this example we study an example of nonmonotonic evolution
in the power spectrum.
Nonmonotonic behaviour can occur for potentials which give rise to
evolution less trivial than a single turn.
We consider
\begin{equation}
V= \sum_{\substack{1 \leq m  \leq 3 \\ 1 \leq n \leq 3}}\left[a_{mn} \cos{\left(m\phi + n\chi\right)}+b_{mn} \sin{\left(m\phi+n\chi\right)}\right].
\label{eq:random-potential}
\end{equation}
The coefficients $a_{mn}$ and $b_{mn}$
are drawn randomly,
making~\label{eq:random-potential}
a low-order Taylor approximation to a ``generic'' potential.
Frazer \& Liddle~\cite{Frazer:2011tg} used potentials of this kind
to study the generic features of inflation in a randomly generated landscape.

Fig.~\ref{nonmono} shows the evolution of the power spectrum and
spectral index in a specific realization of this model,
both of which are nonmonotonic.
Unlike the simple case of double quadratic inflation, the
spectral index does not mirror the evolution of the power spectrum.
This is a consequence of the different initial conditions for
$n_{\alpha\beta}$ and $\Sigma_{\alpha\beta}$,
which make $n_{\alpha\beta}$ grow more slowly than $\Sigma_{\alpha\beta}$.
Eq.~\eqref{eq:ns} implies that
the net result is a decrease in $n_s-1$.

\begin{figure}
\centering
\includegraphics[width=6.2cm]{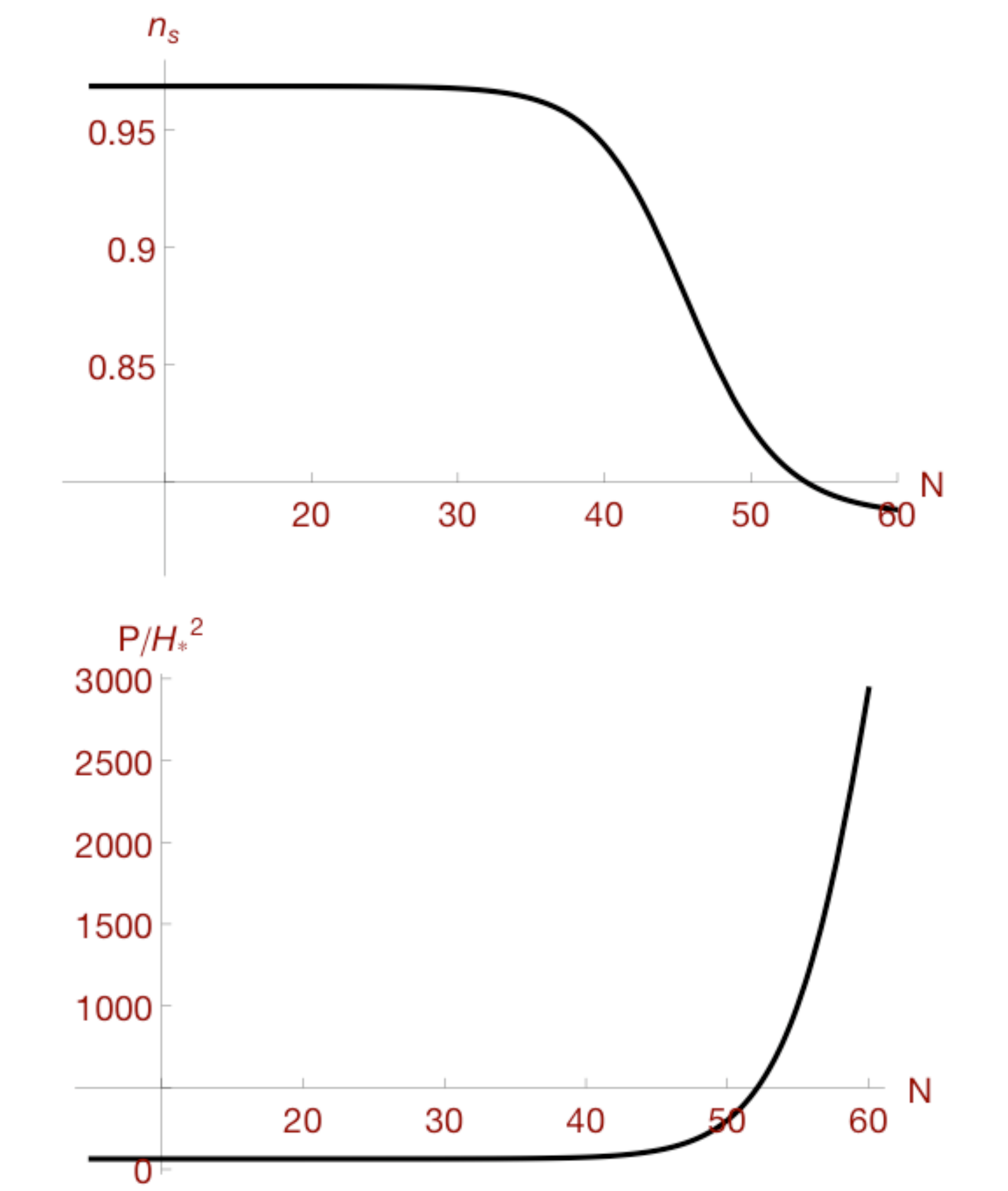}
\caption{Superhorizon evolution of the spectral index (top) and power spectrum of curvature perturbations (bottom) for the exponential quadratic potential.}
\label{chris}
\end{figure}

\para{Exponential quadratic potential}
Our final example exhibits a potential with unusual
characteristics:
the isocurvature modes do not decay
and are present at the end of inflation.
Therefore the curvature perturbation $\zeta$ will continue to
evolve: it has not yet reached its ``adiabatic limit''
\cite{Elliston:2011dr}.
The observable statistical properties of $\zeta$ in models of this type
will depend on unknown details of the post-inflationary
history, including the details of reheating.
Therefore such models are less \emph{predictive} than those which converge
to their adiabatic limit
during inflation.
However,
this is a question of calculability rather than principle.
Models of this kind need not be
intrinsically less interesting
from the point of view of fundamental physics.

We use a model introduced by Byrnes~{\etal}~\cite{Byrnes:2008wi},
and studied in more detail by Elliston~{\etal}~\cite{Elliston:2011dr}.
The potential is
\begin{equation}
V=V_{0}\chi^{2}\e{-\lambda \phi^2} .
\end{equation}
We choose $V_{0}=10^{-5} \Mp^2$ and $\lambda=0.05$,
making our results comparable to Ref.~\cite{Elliston:2011dr} (the
field values 55 e-folds before the end of inflation are $\phi=0.001 \Mp$
and $\chi=16 \Mp$).
Fig.~\ref{chris} shows that
the power grows strongly from roughly 45 e-folds
after horizon crossing, reflecting
the divergence of nearby trajectories in field space
described by Elliston~{\etal}
The corresponding evolution of $n_s$ is a stretched step.
The amplitude of $P$
grows in comparison with
$N_{,\alpha} N_{,\beta} n_{\alpha\beta}$,
and the spectral index decreases significantly.
Its value at the end of inflation is observationally unacceptable,
but nothing can be concluded until the post-inflationary history
is specified.

\section{Summary}

Inflationary models motivated by high-energy physics
frequently involve many light scalar fields
with complicated potentials.
For this reason, it is important to find efficient numerical
tools with which to predict
observable parameters. 

In this paper, we have presented
a simple method to compute the evolution of the spectral index in models
where multi-field dynamics must be taken into account.
Our approach uses a system of ``transport equations''---a set of
ordinary differential equations---to evolve the spectral indices
$n_{\alpha\beta}$
of the 2-point functions of the canonically normalized
scalar fields.
These can be assembled into $n_s$ at the time of interest.

The transport equation for the indices
$n_{\alpha\beta}$ is exactly the same as the equation used to
transport the power spectra of the field perturbations.
However,
owing to its different initial conditions, its evolution
is not trivially related to the evolution of the power spectra.
Its behaviour depends on the characteristics of the
potential.
In~\S\ref{sec:examples} we exhibited
specific examples that illustrate this behaviour.
Although
our examples were drawn from the class of 2-field models for reasons
of simplicity, the
method is valid for any number of fields.

In a model with $N$ fields, we require $N$ integrations to track the
evolution of the centroid $\Phi_\alpha$,
and $N(N+1)/2$ to evolve the two-point function $\Sigma_{\alpha\beta}$.
To obtain a simple estimate,
we assume the computational time is dominated by integration,
and that integrating each component of $\Phi_\alpha$,
$\Sigma_{\alpha\beta}$.
Under these assumptions we can expect the
time required to scale like $N(N+3)/2$.

Our method requires
the introduction of another $N(N+1)/2$ equations to track the evolution of
$n_{\alpha\beta}$.
Assuming integration of $n_{\alpha\beta}$ takes comparable time
to $\Sigma_{\alpha\beta}$ and $\Phi_{\alpha}$ (which
should be a good approximation because $\Sigma_{\alpha\beta}$
and $n_{\alpha\beta}$ obey the same equation),
the total time will now scale like $N(N+2)$.
On the other hand, if we evolve the 2-point function at multiple scales
in order to extract a numerical derivative
by fitting a spline with $m$ points, then we will require
$N + m N(N+1)/2$ integrations.
Typically $m$ will be of order a few.

These simple estimates show that, asymptotically, either method has complexity
$\Or(N^2)$.
%but we might expect the transport method
%to be more efficient by a numerical factor $\sim 2/m$ in the limit
%$N \rightarrow \infty$.
To obtain reasonable accuracy when extracting a numerical derivative,
one might wish to fit
a spline with several points.
In this case one might expect the transport approach to be more
efficient by a modest numerical factor.
%before extracting a numerical derivative,
%making the improvement a factor of $2/3$ as $N \rightarrow \infty$.
%With this choice, at smaller $N$, the improvement is a factor
%typically slightly smaller than $3/4$.
Although not dramatic, a small speed increase
%by a factor in the range $0.66$--$0.75$
would certainly be worthwhile
when accumulated over a large number of samples.

A similar approach could be adopted to calculate the scale dependence of
any $n$-point function. The case of observational interest is likely to
be the bispectrum, where one might wish to compute the running
of the local-mode amplitude $\fNL$
\cite{Byrnes:2010ft, *Byrnes:2010xd}.
We leave this for future work.

%======================================%
%<<<<<<<<<<< ACKNOWLEDGMENTS >>>>>>>>>>%
%======================================%

\begin{acknowledgments}
We would like to thank Jonathan Frazer for discussions,
and providing us with his numerical implementation of
the transport equations.
Andrew Liddle and David Mulryne made helpful comments on an early version of the
manuscript.
M.D.\ was supported by FCT (Portugal).
DS was supported by the Science and Technology Facilities Council
[grant number ST/I000976/1].
\end{acknowledgments}

%======================================%
%<<<<<<<<<<<< BIBLIOGRAPHY >>>>>>>>>>>>%
%======================================%

\bibliography{paper}

%merlin.mbs 2010-03-15 4.21a (PWD, AO, DPC)
\begin{thebibliography}{43}%
\makeatletter
\providecommand \@ifxundefined [1]{%
 \@ifx{#1\undefined}
}%
\providecommand \@ifnum [1]{%
 \ifnum #1\expandafter \@firstoftwo
 \else \expandafter \@secondoftwo
 \fi
}%
\providecommand \@ifx [1]{%
 \ifx #1\expandafter \@firstoftwo
 \else \expandafter \@secondoftwo
 \fi
}%
\providecommand \natexlab [1]{#1}%
\providecommand \enquote  [1]{``#1''}%
\providecommand \bibnamefont  [1]{#1}%
\providecommand \bibfnamefont [1]{#1}%
\providecommand \citenamefont [1]{#1}%
\providecommand \href@noop [0]{\@secondoftwo}%
\providecommand \href [0]{\begingroup \@sanitize@url \@href}%
\providecommand \@href[1]{\@@startlink{#1}\@@href}%
\providecommand \@@href[1]{\endgroup#1\@@endlink}%
\providecommand \@sanitize@url [0]{\catcode `\\12\catcode `\$12\catcode
  `\&12\catcode `\#12\catcode `\^12\catcode `\_12\catcode `\%12\relax}%
\providecommand \@@startlink[1]{}%
\providecommand \@@endlink[0]{}%
\providecommand \url  [0]{\begingroup\@sanitize@url \@url }%
\providecommand \@url [1]{\endgroup\@href {#1}{\urlprefix }}%
\providecommand \urlprefix  [0]{URL }%
\providecommand \Eprint [0]{\href }%
\@ifxundefined \urlstyle {%
  \providecommand \doi  [0]{\begingroup \@sanitize@url \@doi}%
  \providecommand \@doi [1]{\endgroup \@@startlink {\doibase
  #1}doi:\discretionary {}{}{}#1\@@endlink }%
}{%
  \providecommand \doi  [0]{doi:\discretionary{}{}{}\begingroup
  \urlstyle{rm}\Url }%
}%
\providecommand \doibase [0]{http://dx.doi.org/}%
\providecommand \Doi [0]{\begingroup \@sanitize@url \@Doi }%
\providecommand \@Doi  [1]{\endgroup\@@startlink{\doibase#1}\@@Doi}%
\providecommand \@@Doi [1]{#1\@@endlink}%
\providecommand \selectlanguage [0]{\@gobble}%
\providecommand \bibinfo  [0]{\@secondoftwo}%
\providecommand \bibfield  [0]{\@secondoftwo}%
\providecommand \translation [1]{[#1]}%
\providecommand \BibitemOpen [0]{}%
\providecommand \bibitemStop [0]{}%
\providecommand \bibitemNoStop [0]{.\EOS\space}%
\providecommand \EOS [0]{\spacefactor3000\relax}%
\providecommand \BibitemShut  [1]{\csname bibitem#1\endcsname}%
%</preamble>
\bibitem [{\citenamefont {Colombo}\ \emph {et~al.}(2009)\citenamefont
  {Colombo}, \citenamefont {Pierpaoli},\ and\ \citenamefont
  {Pritchard}}]{Colombo:2008ta}%
  \BibitemOpen
  \bibfield  {author} {\bibinfo {author} {\bibfnamefont {L.}~\bibnamefont
  {Colombo}}, \bibinfo {author} {\bibfnamefont {E.}~\bibnamefont {Pierpaoli}},
  \ and\ \bibinfo {author} {\bibfnamefont {J.}~\bibnamefont {Pritchard}},\
  }\Doi {10.1111/j.1365-2966.2009.14802.x} {\bibfield  {journal} {\bibinfo
  {journal} {Mon.Not.Roy.Astron.Soc.},\ }\textbf {\bibinfo {volume} {398}},\
  \bibinfo {pages} {1621} (\bibinfo {year} {2009})},\ \Eprint
  {http://arxiv.org/abs/0811.2622} {arXiv:0811.2622 [astro-ph]} \BibitemShut
  {NoStop}%
\bibitem [{\citenamefont {Starobinsky}(1985)}]{Starobinsky:1986fxa}%
  \BibitemOpen
  \bibfield  {author} {\bibinfo {author} {\bibfnamefont {A.~A.}\ \bibnamefont
  {Starobinsky}},\ }\href@noop {} {\bibfield  {journal} {\bibinfo  {journal}
  {JETP Lett.},\ }\textbf {\bibinfo {volume} {42}},\ \bibinfo {pages} {152}
  (\bibinfo {year} {1985})}\BibitemShut {NoStop}%
\bibitem [{\citenamefont {Lyth}\ and\ \citenamefont
  {Rodr{\'{\i}}guez}(2005)}]{Lyth:2005fi}%
  \BibitemOpen
  \bibfield  {author} {\bibinfo {author} {\bibfnamefont {D.~H.}\ \bibnamefont
  {Lyth}}\ and\ \bibinfo {author} {\bibfnamefont {Y.}~\bibnamefont
  {Rodr{\'{\i}}guez}},\ }\Doi {10.1103/PhysRevLett.95.121302} {\bibfield
  {journal} {\bibinfo  {journal} {Phys.Rev.Lett.},\ }\textbf {\bibinfo {volume}
  {95}},\ \bibinfo {pages} {121302} (\bibinfo {year} {2005})},\ \Eprint
  {http://arxiv.org/abs/astro-ph/0504045} {arXiv:astro-ph/0504045 [astro-ph]}
  \BibitemShut {NoStop}%
\bibitem [{\citenamefont {Sasaki}\ and\ \citenamefont
  {Stewart}(1996)}]{Sasaki:1995aw}%
  \BibitemOpen
  \bibfield  {author} {\bibinfo {author} {\bibfnamefont {M.}~\bibnamefont
  {Sasaki}}\ and\ \bibinfo {author} {\bibfnamefont {E.~D.}\ \bibnamefont
  {Stewart}},\ }\Doi {10.1143/PTP.95.71} {\bibfield  {journal} {\bibinfo
  {journal} {Prog.Theor.Phys.},\ }\textbf {\bibinfo {volume} {95}},\ \bibinfo
  {pages} {71} (\bibinfo {year} {1996})},\ \Eprint
  {http://arxiv.org/abs/astro-ph/9507001} {arXiv:astro-ph/9507001 [astro-ph]}
  \BibitemShut {NoStop}%
\bibitem [{\citenamefont {Frazer}\ and\ \citenamefont
  {Liddle}(2011)}]{Frazer:2011tg}%
  \BibitemOpen
  \bibfield  {author} {\bibinfo {author} {\bibfnamefont {J.}~\bibnamefont
  {Frazer}}\ and\ \bibinfo {author} {\bibfnamefont {A.~R.}\ \bibnamefont
  {Liddle}},\ }\Doi {10.1088/1475-7516/2011/02/026} {\bibfield  {journal}
  {\bibinfo  {journal} {JCAP},\ }\textbf {\bibinfo {volume} {1102}},\ \bibinfo
  {pages} {026} (\bibinfo {year} {2011})},\ \Eprint
  {http://arxiv.org/abs/1101.1619} {arXiv:1101.1619 [astro-ph.CO]} \BibitemShut
  {NoStop}%
\bibitem [{\citenamefont {Agarwal}\ \emph {et~al.}(2011)\citenamefont
  {Agarwal}, \citenamefont {Bean}, \citenamefont {McAllister},\ and\
  \citenamefont {Xu}}]{Agarwal:2011wm}%
  \BibitemOpen
  \bibfield  {author} {\bibinfo {author} {\bibfnamefont {N.}~\bibnamefont
  {Agarwal}}, \bibinfo {author} {\bibfnamefont {R.}~\bibnamefont {Bean}},
  \bibinfo {author} {\bibfnamefont {L.}~\bibnamefont {McAllister}}, \ and\
  \bibinfo {author} {\bibfnamefont {G.}~\bibnamefont {Xu}},\ }\Doi
  {10.1088/1475-7516/2011/09/002} {\bibfield  {journal} {\bibinfo  {journal}
  {JCAP},\ }\textbf {\bibinfo {volume} {1109}},\ \bibinfo {pages} {002}
  (\bibinfo {year} {2011})},\ \Eprint {http://arxiv.org/abs/1103.2775}
  {arXiv:1103.2775 [astro-ph.CO]} \BibitemShut {NoStop}%
\bibitem [{\citenamefont {Mulryne}\ \emph {et~al.}(2011)\citenamefont
  {Mulryne}, \citenamefont {Seery},\ and\ \citenamefont
  {Wesley}}]{Mulryne:2010rp}%
  \BibitemOpen
  \bibfield  {author} {\bibinfo {author} {\bibfnamefont {D.~J.}\ \bibnamefont
  {Mulryne}}, \bibinfo {author} {\bibfnamefont {D.}~\bibnamefont {Seery}}, \
  and\ \bibinfo {author} {\bibfnamefont {D.}~\bibnamefont {Wesley}},\ }\Doi
  {10.1088/1475-7516/2011/04/030} {\bibfield  {journal} {\bibinfo  {journal}
  {JCAP},\ }\textbf {\bibinfo {volume} {1104}},\ \bibinfo {pages} {030}
  (\bibinfo {year} {2011})},\ \Eprint {http://arxiv.org/abs/1008.3159}
  {arXiv:1008.3159 [astro-ph.CO]} \BibitemShut {NoStop}%
\bibitem [{\citenamefont {Gordon}\ \emph {et~al.}(2001)\citenamefont {Gordon},
  \citenamefont {Wands}, \citenamefont {Bassett},\ and\ \citenamefont
  {Maartens}}]{Gordon:2000hv}%
  \BibitemOpen
  \bibfield  {author} {\bibinfo {author} {\bibfnamefont {C.}~\bibnamefont
  {Gordon}}, \bibinfo {author} {\bibfnamefont {D.}~\bibnamefont {Wands}},
  \bibinfo {author} {\bibfnamefont {B.~A.}\ \bibnamefont {Bassett}}, \ and\
  \bibinfo {author} {\bibfnamefont {R.}~\bibnamefont {Maartens}},\ }\Doi
  {10.1103/PhysRevD.63.023506} {\bibfield  {journal} {\bibinfo  {journal}
  {Phys.Rev.},\ }\textbf {\bibinfo {volume} {D63}},\ \bibinfo {pages} {023506}
  (\bibinfo {year} {2001})},\ \Eprint {http://arxiv.org/abs/astro-ph/0009131}
  {arXiv:astro-ph/0009131 [astro-ph]} \BibitemShut {NoStop}%
\bibitem [{\citenamefont {Groot~Nibbelink}\ and\ \citenamefont {van
  Tent}(2002)}]{GrootNibbelink:2001qt}%
  \BibitemOpen
  \bibfield  {author} {\bibinfo {author} {\bibfnamefont {S.}~\bibnamefont
  {Groot~Nibbelink}}\ and\ \bibinfo {author} {\bibfnamefont {B.}~\bibnamefont
  {van Tent}},\ }\Doi {10.1088/0264-9381/19/4/302} {\bibfield  {journal}
  {\bibinfo  {journal} {Class.Quant.Grav.},\ }\textbf {\bibinfo {volume}
  {19}},\ \bibinfo {pages} {613} (\bibinfo {year} {2002})},\ \Eprint
  {http://arxiv.org/abs/hep-ph/0107272} {arXiv:hep-ph/0107272 [hep-ph]}
  \BibitemShut {NoStop}%
\bibitem [{\citenamefont {Peterson}\ and\ \citenamefont
  {Tegmark}(2011){\natexlab{a}}}]{Peterson:2010np}%
  \BibitemOpen
  \bibfield  {author} {\bibinfo {author} {\bibfnamefont {C.~M.}\ \bibnamefont
  {Peterson}}\ and\ \bibinfo {author} {\bibfnamefont {M.}~\bibnamefont
  {Tegmark}},\ }\Doi {10.1103/PhysRevD.83.023522} {\bibfield  {journal}
  {\bibinfo  {journal} {Phys.Rev.},\ }\textbf {\bibinfo {volume} {D83}},\
  \bibinfo {pages} {023522} (\bibinfo {year} {2011}{\natexlab{a}})},\ \Eprint
  {http://arxiv.org/abs/1005.4056} {arXiv:1005.4056 [astro-ph.CO]} \BibitemShut
  {NoStop}%
\bibitem [{\citenamefont {Peterson}\ and\ \citenamefont
  {Tegmark}(2011){\natexlab{b}}}]{Peterson:2010mv}%
  \BibitemOpen
  \bibfield  {author} {\bibinfo {author} {\bibfnamefont {C.~M.}\ \bibnamefont
  {Peterson}}\ and\ \bibinfo {author} {\bibfnamefont {M.}~\bibnamefont
  {Tegmark}},\ }\Doi {10.1103/PhysRevD.84.023520} {\bibfield  {journal}
  {\bibinfo  {journal} {Phys.Rev.},\ }\textbf {\bibinfo {volume} {D84}},\
  \bibinfo {pages} {023520} (\bibinfo {year} {2011}{\natexlab{b}})},\ \Eprint
  {http://arxiv.org/abs/1011.6675} {arXiv:1011.6675 [astro-ph.CO]} \BibitemShut
  {NoStop}%
\bibitem [{\citenamefont {Peterson}\ and\ \citenamefont
  {Tegmark}(2011){\natexlab{c}}}]{Peterson:2011yt}%
  \BibitemOpen
  \bibfield  {author} {\bibinfo {author} {\bibfnamefont {C.~M.}\ \bibnamefont
  {Peterson}}\ and\ \bibinfo {author} {\bibfnamefont {M.}~\bibnamefont
  {Tegmark}},\ }\href@noop {} { (\bibinfo {year} {2011}{\natexlab{c}})},\
  \Eprint {http://arxiv.org/abs/1111.0927} {arXiv:1111.0927 [astro-ph.CO]}
  \BibitemShut {NoStop}%
\bibitem [{\citenamefont {Mulryne}\ \emph {et~al.}(2010)\citenamefont
  {Mulryne}, \citenamefont {Seery},\ and\ \citenamefont
  {Wesley}}]{Mulryne:2009kh}%
  \BibitemOpen
  \bibfield  {author} {\bibinfo {author} {\bibfnamefont {D.~J.}\ \bibnamefont
  {Mulryne}}, \bibinfo {author} {\bibfnamefont {D.}~\bibnamefont {Seery}}, \
  and\ \bibinfo {author} {\bibfnamefont {D.}~\bibnamefont {Wesley}},\ }\Doi
  {10.1088/1475-7516/2010/01/024} {\bibfield  {journal} {\bibinfo  {journal}
  {JCAP},\ }\textbf {\bibinfo {volume} {1001}},\ \bibinfo {pages} {024}
  (\bibinfo {year} {2010})},\ \Eprint {http://arxiv.org/abs/0909.2256}
  {arXiv:0909.2256 [astro-ph.CO]} \BibitemShut {NoStop}%
\bibitem [{\citenamefont {Lyth}(1985)}]{Lyth:1984gv}%
  \BibitemOpen
  \bibfield  {author} {\bibinfo {author} {\bibfnamefont {D.}~\bibnamefont
  {Lyth}},\ }\Doi {10.1103/PhysRevD.31.1792} {\bibfield  {journal} {\bibinfo
  {journal} {Phys.Rev.},\ }\textbf {\bibinfo {volume} {D31}},\ \bibinfo {pages}
  {1792} (\bibinfo {year} {1985})}\BibitemShut {NoStop}%
\bibitem [{\citenamefont {Polarski}\ and\ \citenamefont
  {Starobinsky}(1996)}]{Polarski:1995jg}%
  \BibitemOpen
  \bibfield  {author} {\bibinfo {author} {\bibfnamefont {D.}~\bibnamefont
  {Polarski}}\ and\ \bibinfo {author} {\bibfnamefont {A.~A.}\ \bibnamefont
  {Starobinsky}},\ }\Doi {10.1088/0264-9381/13/3/006} {\bibfield  {journal}
  {\bibinfo  {journal} {Class.Quant.Grav.},\ }\textbf {\bibinfo {volume}
  {13}},\ \bibinfo {pages} {377} (\bibinfo {year} {1996})},\ \Eprint
  {http://arxiv.org/abs/gr-qc/9504030} {arXiv:gr-qc/9504030 [gr-qc]}
  \BibitemShut {NoStop}%
\bibitem [{\citenamefont {Lyth}\ and\ \citenamefont
  {Seery}(2008)}]{Lyth:2006qz}%
  \BibitemOpen
  \bibfield  {author} {\bibinfo {author} {\bibfnamefont {D.~H.}\ \bibnamefont
  {Lyth}}\ and\ \bibinfo {author} {\bibfnamefont {D.}~\bibnamefont {Seery}},\
  }\Doi {10.1016/j.physletb.2008.03.010} {\bibfield  {journal} {\bibinfo
  {journal} {Phys.Lett.},\ }\textbf {\bibinfo {volume} {B662}},\ \bibinfo
  {pages} {309} (\bibinfo {year} {2008})},\ \Eprint
  {http://arxiv.org/abs/astro-ph/0607647} {arXiv:astro-ph/0607647 [astro-ph]}
  \BibitemShut {NoStop}%
\bibitem [{\citenamefont {Gardiner}(2002)}]{citeulike:1400625}%
  \BibitemOpen
  \bibfield  {author} {\bibinfo {author} {\bibfnamefont {C.~W.}\ \bibnamefont
  {Gardiner}},\ }\href@noop {} {\emph {\bibinfo {title} {{Handbook of
  stochastic methods: for physics, chemistry and the natural sciences}}}},\
  Springer Series in Synergetics, 13\ (\bibinfo  {publisher} {Springer},\
  \bibinfo {year} {2002})\ ISBN \bibinfo {isbn} {3540616349}\BibitemShut
  {NoStop}%
\bibitem [{\citenamefont {Calzetta}\ and\ \citenamefont
  {Hu}(1997)}]{Calzetta:1996sy}%
  \BibitemOpen
  \bibfield  {author} {\bibinfo {author} {\bibfnamefont {E.}~\bibnamefont
  {Calzetta}}\ and\ \bibinfo {author} {\bibfnamefont {B.-L.}\ \bibnamefont
  {Hu}},\ }\Doi {10.1103/PhysRevD.55.3536} {\bibfield  {journal} {\bibinfo
  {journal} {Phys.Rev.},\ }\textbf {\bibinfo {volume} {D55}},\ \bibinfo {pages}
  {3536} (\bibinfo {year} {1997})},\ \Eprint
  {http://arxiv.org/abs/hep-th/9603164} {arXiv:hep-th/9603164 [hep-th]}
  \BibitemShut {NoStop}%
\bibitem [{\citenamefont {Starobinsky}(1986)}]{Starobinsky:1986fx}%
  \BibitemOpen
  \bibfield  {author} {\bibinfo {author} {\bibfnamefont {A.~A.}\ \bibnamefont
  {Starobinsky}},\ }in\ \href@noop {} {\emph {\bibinfo {booktitle} {Field
  Theory, Quantum Gravity and Strings}}},\ \bibinfo {editor} {edited by\
  \bibinfo {editor} {\bibfnamefont {H.}~\bibnamefont {De~Vega}}\ and\ \bibinfo
  {editor} {\bibfnamefont {N.}~\bibnamefont {Sanchez}}}\ (\bibinfo {year}
  {1986})\ pp.\ \bibinfo {pages} {107--126}\BibitemShut {NoStop}%
\bibitem [{\citenamefont {Starobinsky}\ and\ \citenamefont
  {Yokoyama}(1994)}]{Starobinsky:1994bd}%
  \BibitemOpen
  \bibfield  {author} {\bibinfo {author} {\bibfnamefont {A.~A.}\ \bibnamefont
  {Starobinsky}}\ and\ \bibinfo {author} {\bibfnamefont {J.}~\bibnamefont
  {Yokoyama}},\ }\Doi {10.1103/PhysRevD.50.6357} {\bibfield  {journal}
  {\bibinfo  {journal} {Phys.Rev.},\ }\textbf {\bibinfo {volume} {D50}},\
  \bibinfo {pages} {6357} (\bibinfo {year} {1994})},\ \Eprint
  {http://arxiv.org/abs/astro-ph/9407016} {arXiv:astro-ph/9407016 [astro-ph]}
  \BibitemShut {NoStop}%
\bibitem [{\citenamefont {Salopek}\ and\ \citenamefont
  {Bond}(1991)}]{Salopek:1990re}%
  \BibitemOpen
  \bibfield  {author} {\bibinfo {author} {\bibfnamefont {D.}~\bibnamefont
  {Salopek}}\ and\ \bibinfo {author} {\bibfnamefont {J.}~\bibnamefont {Bond}},\
  }\Doi {10.1103/PhysRevD.43.1005} {\bibfield  {journal} {\bibinfo  {journal}
  {Phys.Rev.},\ }\textbf {\bibinfo {volume} {D43}},\ \bibinfo {pages} {1005}
  (\bibinfo {year} {1991})}\BibitemShut {NoStop}%
\bibitem [{\citenamefont {Salopek}\ and\ \citenamefont
  {Bond}(1990)}]{Salopek:1990jq}%
  \BibitemOpen
  \bibfield  {author} {\bibinfo {author} {\bibfnamefont {D.}~\bibnamefont
  {Salopek}}\ and\ \bibinfo {author} {\bibfnamefont {J.}~\bibnamefont {Bond}},\
  }\Doi {10.1103/PhysRevD.42.3936} {\bibfield  {journal} {\bibinfo  {journal}
  {Phys.Rev.},\ }\textbf {\bibinfo {volume} {D42}},\ \bibinfo {pages} {3936}
  (\bibinfo {year} {1990})}\BibitemShut {NoStop}%
\bibitem [{\citenamefont {Seery}(2009)}]{Seery:2009hs}%
  \BibitemOpen
  \bibfield  {author} {\bibinfo {author} {\bibfnamefont {D.}~\bibnamefont
  {Seery}},\ }\Doi {10.1088/1475-7516/2009/05/021} {\bibfield  {journal}
  {\bibinfo  {journal} {JCAP},\ }\textbf {\bibinfo {volume} {0905}},\ \bibinfo
  {pages} {021} (\bibinfo {year} {2009})},\ \Eprint
  {http://arxiv.org/abs/0903.2788} {arXiv:0903.2788 [astro-ph.CO]} \BibitemShut
  {NoStop}%
\bibitem [{\citenamefont {Calzetta}\ and\ \citenamefont
  {Hu}(2000)}]{Calzetta:1999xh}%
  \BibitemOpen
  \bibfield  {author} {\bibinfo {author} {\bibfnamefont {E.}~\bibnamefont
  {Calzetta}}\ and\ \bibinfo {author} {\bibfnamefont {B.}~\bibnamefont {Hu}},\
  }\Doi {10.1103/PhysRevD.61.025012} {\bibfield  {journal} {\bibinfo  {journal}
  {Phys.Rev.},\ }\textbf {\bibinfo {volume} {D61}},\ \bibinfo {pages} {025012}
  (\bibinfo {year} {2000})},\ \Eprint {http://arxiv.org/abs/hep-ph/9903291}
  {arXiv:hep-ph/9903291 [hep-ph]} \BibitemShut {NoStop}%
\bibitem [{\citenamefont {Boubekeur}\ and\ \citenamefont
  {Lyth}(2006)}]{Boubekeur:2005fj}%
  \BibitemOpen
  \bibfield  {author} {\bibinfo {author} {\bibfnamefont {L.}~\bibnamefont
  {Boubekeur}}\ and\ \bibinfo {author} {\bibfnamefont {D.}~\bibnamefont
  {Lyth}},\ }\Doi {10.1103/PhysRevD.73.021301} {\bibfield  {journal} {\bibinfo
  {journal} {Phys.Rev.},\ }\textbf {\bibinfo {volume} {D73}},\ \bibinfo {pages}
  {021301} (\bibinfo {year} {2006})},\ \Eprint
  {http://arxiv.org/abs/astro-ph/0504046} {arXiv:astro-ph/0504046 [astro-ph]}
  \BibitemShut {NoStop}%
\bibitem [{\citenamefont {Lyth}(2007)}]{Lyth:2007jh}%
  \BibitemOpen
  \bibfield  {author} {\bibinfo {author} {\bibfnamefont {D.~H.}\ \bibnamefont
  {Lyth}},\ }\Doi {10.1088/1475-7516/2007/12/016} {\bibfield  {journal}
  {\bibinfo  {journal} {JCAP},\ }\textbf {\bibinfo {volume} {0712}},\ \bibinfo
  {pages} {016} (\bibinfo {year} {2007})},\ \Eprint
  {http://arxiv.org/abs/0707.0361} {arXiv:0707.0361 [astro-ph]} \BibitemShut
  {NoStop}%
\bibitem [{\citenamefont {Seery}(2008)}]{Seery:2007wf}%
  \BibitemOpen
  \bibfield  {author} {\bibinfo {author} {\bibfnamefont {D.}~\bibnamefont
  {Seery}},\ }\Doi {10.1088/1475-7516/2008/02/006} {\bibfield  {journal}
  {\bibinfo  {journal} {JCAP},\ }\textbf {\bibinfo {volume} {0802}},\ \bibinfo
  {pages} {006} (\bibinfo {year} {2008})},\ \Eprint
  {http://arxiv.org/abs/0707.3378} {arXiv:0707.3378 [astro-ph]} \BibitemShut
  {NoStop}%
\bibitem [{\citenamefont {Bartolo}\ \emph {et~al.}(2008)\citenamefont
  {Bartolo}, \citenamefont {Matarrese}, \citenamefont {Pietroni}, \citenamefont
  {Riotto},\ and\ \citenamefont {Seery}}]{Bartolo:2007ti}%
  \BibitemOpen
  \bibfield  {author} {\bibinfo {author} {\bibfnamefont {N.}~\bibnamefont
  {Bartolo}}, \bibinfo {author} {\bibfnamefont {S.}~\bibnamefont {Matarrese}},
  \bibinfo {author} {\bibfnamefont {M.}~\bibnamefont {Pietroni}}, \bibinfo
  {author} {\bibfnamefont {A.}~\bibnamefont {Riotto}}, \ and\ \bibinfo {author}
  {\bibfnamefont {D.}~\bibnamefont {Seery}},\ }\Doi
  {10.1088/1475-7516/2008/01/015} {\bibfield  {journal} {\bibinfo  {journal}
  {JCAP},\ }\textbf {\bibinfo {volume} {0801}},\ \bibinfo {pages} {015}
  (\bibinfo {year} {2008})},\ \Eprint {http://arxiv.org/abs/0711.4263}
  {arXiv:0711.4263 [astro-ph]} \BibitemShut {NoStop}%
\bibitem [{\citenamefont {Seery}(2010)}]{Seery:2010kh}%
  \BibitemOpen
  \bibfield  {author} {\bibinfo {author} {\bibfnamefont {D.}~\bibnamefont
  {Seery}},\ }\Doi {10.1088/0264-9381/27/12/124005} {\bibfield  {journal}
  {\bibinfo  {journal} {Class.Quant.Grav.},\ }\textbf {\bibinfo {volume}
  {27}},\ \bibinfo {pages} {124005} (\bibinfo {year} {2010})},\ \Eprint
  {http://arxiv.org/abs/1005.1649} {arXiv:1005.1649 [astro-ph.CO]} \BibitemShut
  {NoStop}%
\bibitem [{\citenamefont {Bardeen}(1980)}]{Bardeen:1980kt}%
  \BibitemOpen
  \bibfield  {author} {\bibinfo {author} {\bibfnamefont {J.~M.}\ \bibnamefont
  {Bardeen}},\ }\Doi {10.1103/PhysRevD.22.1882} {\bibfield  {journal} {\bibinfo
   {journal} {Phys.Rev.},\ }\textbf {\bibinfo {volume} {D22}},\ \bibinfo
  {pages} {1882} (\bibinfo {year} {1980})}\BibitemShut {NoStop}%
\bibitem [{\citenamefont {Bardeen}\ \emph {et~al.}(1983)\citenamefont
  {Bardeen}, \citenamefont {Steinhardt},\ and\ \citenamefont
  {Turner}}]{Bardeen:1983qw}%
  \BibitemOpen
  \bibfield  {author} {\bibinfo {author} {\bibfnamefont {J.~M.}\ \bibnamefont
  {Bardeen}}, \bibinfo {author} {\bibfnamefont {P.~J.}\ \bibnamefont
  {Steinhardt}}, \ and\ \bibinfo {author} {\bibfnamefont {M.~S.}\ \bibnamefont
  {Turner}},\ }\Doi {10.1103/PhysRevD.28.679} {\bibfield  {journal} {\bibinfo
  {journal} {Phys.Rev.},\ }\textbf {\bibinfo {volume} {D28}},\ \bibinfo {pages}
  {679} (\bibinfo {year} {1983})}\BibitemShut {NoStop}%
\bibitem [{\citenamefont {Wands}\ \emph {et~al.}(2000)\citenamefont {Wands},
  \citenamefont {Malik}, \citenamefont {Lyth},\ and\ \citenamefont
  {Liddle}}]{Wands:2000dp}%
  \BibitemOpen
  \bibfield  {author} {\bibinfo {author} {\bibfnamefont {D.}~\bibnamefont
  {Wands}}, \bibinfo {author} {\bibfnamefont {K.~A.}\ \bibnamefont {Malik}},
  \bibinfo {author} {\bibfnamefont {D.~H.}\ \bibnamefont {Lyth}}, \ and\
  \bibinfo {author} {\bibfnamefont {A.~R.}\ \bibnamefont {Liddle}},\ }\Doi
  {10.1103/PhysRevD.62.043527} {\bibfield  {journal} {\bibinfo  {journal}
  {Phys.Rev.},\ }\textbf {\bibinfo {volume} {D62}},\ \bibinfo {pages} {043527}
  (\bibinfo {year} {2000})},\ \Eprint {http://arxiv.org/abs/astro-ph/0003278}
  {arXiv:astro-ph/0003278 [astro-ph]} \BibitemShut {NoStop}%
\bibitem [{\citenamefont {Maldacena}(2003)}]{Maldacena:2002vr}%
  \BibitemOpen
  \bibfield  {author} {\bibinfo {author} {\bibfnamefont {J.~M.}\ \bibnamefont
  {Maldacena}},\ }\href@noop {} {\bibfield  {journal} {\bibinfo  {journal}
  {JHEP},\ }\textbf {\bibinfo {volume} {0305}},\ \bibinfo {pages} {013}
  (\bibinfo {year} {2003})},\ \Eprint {http://arxiv.org/abs/astro-ph/0210603}
  {arXiv:astro-ph/0210603 [astro-ph]} \BibitemShut {NoStop}%
\bibitem [{\citenamefont {Malik}\ and\ \citenamefont
  {Wands}(2009)}]{Malik:2008im}%
  \BibitemOpen
  \bibfield  {author} {\bibinfo {author} {\bibfnamefont {K.~A.}\ \bibnamefont
  {Malik}}\ and\ \bibinfo {author} {\bibfnamefont {D.}~\bibnamefont {Wands}},\
  }\Doi {10.1016/j.physrep.2009.03.001} {\bibfield  {journal} {\bibinfo
  {journal} {Phys.Rept.},\ }\textbf {\bibinfo {volume} {475}},\ \bibinfo
  {pages} {1} (\bibinfo {year} {2009})},\ \Eprint
  {http://arxiv.org/abs/0809.4944} {arXiv:0809.4944 [astro-ph]} \BibitemShut
  {NoStop}%
\bibitem [{\citenamefont {Liddle}\ and\ \citenamefont
  {Lyth}(1992)}]{Liddle:1992wi}%
  \BibitemOpen
  \bibfield  {author} {\bibinfo {author} {\bibfnamefont {A.~R.}\ \bibnamefont
  {Liddle}}\ and\ \bibinfo {author} {\bibfnamefont {D.~H.}\ \bibnamefont
  {Lyth}},\ }\Doi {10.1016/0370-2693(92)91393-N} {\bibfield  {journal}
  {\bibinfo  {journal} {Phys.Lett.},\ }\textbf {\bibinfo {volume} {B291}},\
  \bibinfo {pages} {391} (\bibinfo {year} {1992})},\ \Eprint
  {http://arxiv.org/abs/astro-ph/9208007} {arXiv:astro-ph/9208007 [astro-ph]}
  \BibitemShut {NoStop}%
\bibitem [{\citenamefont {Langlois}(1999)}]{Langlois:1999dw}%
  \BibitemOpen
  \bibfield  {author} {\bibinfo {author} {\bibfnamefont {D.}~\bibnamefont
  {Langlois}},\ }\Doi {10.1103/PhysRevD.59.123512} {\bibfield  {journal}
  {\bibinfo  {journal} {Phys.Rev.},\ }\textbf {\bibinfo {volume} {D59}},\
  \bibinfo {pages} {123512} (\bibinfo {year} {1999})},\ \Eprint
  {http://arxiv.org/abs/astro-ph/9906080} {arXiv:astro-ph/9906080 [astro-ph]}
  \BibitemShut {NoStop}%
\bibitem [{\citenamefont {Rigopoulos}\ \emph {et~al.}(2006)\citenamefont
  {Rigopoulos}, \citenamefont {Shellard},\ and\ \citenamefont {van
  Tent}}]{Rigopoulos:2005ae}%
  \BibitemOpen
  \bibfield  {author} {\bibinfo {author} {\bibfnamefont {G.}~\bibnamefont
  {Rigopoulos}}, \bibinfo {author} {\bibfnamefont {E.}~\bibnamefont
  {Shellard}}, \ and\ \bibinfo {author} {\bibfnamefont {B.}~\bibnamefont {van
  Tent}},\ }\Doi {10.1103/PhysRevD.73.083522} {\bibfield  {journal} {\bibinfo
  {journal} {Phys.Rev.},\ }\textbf {\bibinfo {volume} {D73}},\ \bibinfo {pages}
  {083522} (\bibinfo {year} {2006})},\ \Eprint
  {http://arxiv.org/abs/astro-ph/0506704} {arXiv:astro-ph/0506704 [astro-ph]}
  \BibitemShut {NoStop}%
\bibitem [{\citenamefont {Rigopoulos}\ \emph {et~al.}(2007)\citenamefont
  {Rigopoulos}, \citenamefont {Shellard},\ and\ \citenamefont {van
  Tent}}]{Rigopoulos:2005us}%
  \BibitemOpen
  \bibfield  {author} {\bibinfo {author} {\bibfnamefont {G.}~\bibnamefont
  {Rigopoulos}}, \bibinfo {author} {\bibfnamefont {E.}~\bibnamefont
  {Shellard}}, \ and\ \bibinfo {author} {\bibfnamefont {B.}~\bibnamefont {van
  Tent}},\ }\Doi {10.1103/PhysRevD.76.083512} {\bibfield  {journal} {\bibinfo
  {journal} {Phys.Rev.},\ }\textbf {\bibinfo {volume} {D76}},\ \bibinfo {pages}
  {083512} (\bibinfo {year} {2007})},\ \Eprint
  {http://arxiv.org/abs/astro-ph/0511041} {arXiv:astro-ph/0511041 [astro-ph]}
  \BibitemShut {NoStop}%
\bibitem [{\citenamefont {Vernizzi}\ and\ \citenamefont
  {Wands}(2006)}]{Vernizzi:2006ve}%
  \BibitemOpen
  \bibfield  {author} {\bibinfo {author} {\bibfnamefont {F.}~\bibnamefont
  {Vernizzi}}\ and\ \bibinfo {author} {\bibfnamefont {D.}~\bibnamefont
  {Wands}},\ }\Doi {10.1088/1475-7516/2006/05/019} {\bibfield  {journal}
  {\bibinfo  {journal} {JCAP},\ }\textbf {\bibinfo {volume} {0605}},\ \bibinfo
  {pages} {019} (\bibinfo {year} {2006})},\ \Eprint
  {http://arxiv.org/abs/astro-ph/0603799} {arXiv:astro-ph/0603799 [astro-ph]}
  \BibitemShut {NoStop}%
\bibitem [{\citenamefont {Elliston}\ \emph {et~al.}(2011)\citenamefont
  {Elliston}, \citenamefont {Mulryne}, \citenamefont {Seery},\ and\
  \citenamefont {Tavakol}}]{Elliston:2011dr}%
  \BibitemOpen
  \bibfield  {author} {\bibinfo {author} {\bibfnamefont {J.}~\bibnamefont
  {Elliston}}, \bibinfo {author} {\bibfnamefont {D.~J.}\ \bibnamefont
  {Mulryne}}, \bibinfo {author} {\bibfnamefont {D.}~\bibnamefont {Seery}}, \
  and\ \bibinfo {author} {\bibfnamefont {R.}~\bibnamefont {Tavakol}},\ }\Doi
  {10.1088/1475-7516/2011/11/005} {\bibfield  {journal} {\bibinfo  {journal}
  {JCAP},\ }\textbf {\bibinfo {volume} {1111}},\ \bibinfo {pages} {005}
  (\bibinfo {year} {2011})},\ \Eprint {http://arxiv.org/abs/1106.2153}
  {arXiv:1106.2153 [astro-ph.CO]} \BibitemShut {NoStop}%
\bibitem [{\citenamefont {Byrnes}\ \emph {et~al.}(2008)\citenamefont {Byrnes},
  \citenamefont {Choi},\ and\ \citenamefont {Hall}}]{Byrnes:2008wi}%
  \BibitemOpen
  \bibfield  {author} {\bibinfo {author} {\bibfnamefont {C.~T.}\ \bibnamefont
  {Byrnes}}, \bibinfo {author} {\bibfnamefont {K.-Y.}\ \bibnamefont {Choi}}, \
  and\ \bibinfo {author} {\bibfnamefont {L.~M.}\ \bibnamefont {Hall}},\ }\Doi
  {10.1088/1475-7516/2008/10/008} {\bibfield  {journal} {\bibinfo  {journal}
  {JCAP},\ }\textbf {\bibinfo {volume} {0810}},\ \bibinfo {pages} {008}
  (\bibinfo {year} {2008})},\ \Eprint {http://arxiv.org/abs/0807.1101}
  {arXiv:0807.1101 [astro-ph]} \BibitemShut {NoStop}%
\bibitem [{\citenamefont {Byrnes}\ \emph
  {et~al.}(2010){\natexlab{a}}\citenamefont {Byrnes}, \citenamefont
  {Gerstenlauer}, \citenamefont {Nurmi}, \citenamefont {Tasinato},\ and\
  \citenamefont {Wands}}]{Byrnes:2010ft}%
  \BibitemOpen
  \bibfield  {author} {\bibinfo {author} {\bibfnamefont {C.~T.}\ \bibnamefont
  {Byrnes}}, \bibinfo {author} {\bibfnamefont {M.}~\bibnamefont
  {Gerstenlauer}}, \bibinfo {author} {\bibfnamefont {S.}~\bibnamefont {Nurmi}},
  \bibinfo {author} {\bibfnamefont {G.}~\bibnamefont {Tasinato}}, \ and\
  \bibinfo {author} {\bibfnamefont {D.}~\bibnamefont {Wands}},\ }\Doi
  {10.1088/1475-7516/2010/10/004} {\bibfield  {journal} {\bibinfo  {journal}
  {JCAP},\ }\textbf {\bibinfo {volume} {1010}},\ \bibinfo {pages} {004}
  (\bibinfo {year} {2010}{\natexlab{a}})},\ \Eprint
  {http://arxiv.org/abs/1007.4277} {arXiv:1007.4277 [astro-ph.CO]} \BibitemShut
  {NoStop}%
\bibitem [{\citenamefont {Byrnes}\ \emph
  {et~al.}(2010){\natexlab{b}}\citenamefont {Byrnes}, \citenamefont {Enqvist},\
  and\ \citenamefont {Takahashi}}]{Byrnes:2010xd}%
  \BibitemOpen
  \bibfield  {author} {\bibinfo {author} {\bibfnamefont {C.~T.}\ \bibnamefont
  {Byrnes}}, \bibinfo {author} {\bibfnamefont {K.}~\bibnamefont {Enqvist}}, \
  and\ \bibinfo {author} {\bibfnamefont {T.}~\bibnamefont {Takahashi}},\ }\Doi
  {10.1088/1475-7516/2010/09/026} {\bibfield  {journal} {\bibinfo  {journal}
  {JCAP},\ }\textbf {\bibinfo {volume} {1009}},\ \bibinfo {pages} {026}
  (\bibinfo {year} {2010}{\natexlab{b}})},\ \Eprint
  {http://arxiv.org/abs/1007.5148} {arXiv:1007.5148 [astro-ph.CO]} \BibitemShut
  {NoStop}%
\end{thebibliography}%

%%%%%%%%%%%%%%%%%%%%%%%%%%%%%%%%%%%%%%%%%%%%%%%%%%%%%%%%%%%%%%%%%%%%%%%%
\end{document}